\documentclass[aoas,nameyear,MSNbibl,seceqn,dvips]{arximspdf}
\usepackage{dcolumn}
\usepackage{graphicx}

\doi{10.1214/11-AOAS456}
\volume{5}
\issue{3}
\pubyear{2011}
\firstpage{1726}
\lastpage{1751}

\makeatletter
\newproclaim{remark}{Remark}
\newproclaim{result}{Result}
\newproclaim{definition}{Definition}
\newtheorem{lemma}{Lemma}
\newcolumntype{d}[1]{D{.}{.}{#1}}

\def\bptnote#1{}

\def\@bmisc[#1]{%
  \get@battribute{unstr}%
  \common@pub@types%
  \let\bauthor\bbl@bauthor%
  \let\bhowpublished\@firstofone%
  \def\borganization##1{{\bauthor@style ##1}}%
}
\makeatother

\begin{document}
\begin{frontmatter}

\title{Are private schools better than public schools? Appraisal for Ireland by methods for observational~studies\protect\thanksref{TITL1}}
\pdftitle{Are private schools better than public schools? Appraisal for Ireland by methods for observational~studies}
\runtitle{Are private schools better than public schools?}

\thankstext{TITL1}{Supported by the Israel Science Foundation Grant 1277/05.}

\begin{aug}
\author[A]{\fnms{Danny} \snm{Pfeffermann}\corref{}\ead[label=e1]{msdanny@soton.ac.uk}}
and
\author[B]{\fnms{Victoria} \snm{Landsman}\ead[label=e2]{landsmanv@mail.nih.gov}}
\runauthor{D. Pfeffermann and V. Landsman}
\affiliation{Hebrew University of Jerusalem and University of Southampton, and~National~Cancer Institute}
\address[A]{Hebrew University of Jerusalem\\
Jerusalem, 91905\\
Israel\\
and\\
University of Southampton\\
Southampton, SO17 1BJ\\
United Kingdom\\
\printead{e1}} 
\address[B]{National Cancer Institute\\
NIH\\
Rockville, Maryland 20852\\
USA\\
\printead{e2}}
\end{aug}

\received{\smonth{8} \syear{2010}}
\revised{\smonth{12} \syear{2010}}

\begin{abstract}
In observational studies the assignment of units to treatments is not
under control. Consequently, the estimation and comparison of treatment
effects based on the empirical distribution of the responses can be
biased since the units exposed to the various treatments could differ in
important unknown pretreatment characteristics, which are related to the
response. An important example studied in this article is the question
of whether private schools offer better quality of education than public
schools. In order to address this question, we use data collected in the
year 2000 by OECD for the \textit{Programme for International Student
Assessment} (PISA). Focusing for illustration on scores in mathematics
of 15-year-old pupils in Ireland, we find that the raw average score of
pupils in private schools is higher than of pupils in public schools.
However, application of a newly proposed method for observational
studies suggests that the less able pupils tend to enroll in public
schools, such that their lower scores are not necessarily an indication
of bad quality of the public schools. Indeed, when comparing the average
score in the two types of schools after adjusting for the enrollment
effects, we find quite surprisingly that public schools perform better
on average. This outcome is supported by the methods of instrumental
variables and latent variables, commonly used by econometricians for
analyzing and evaluating social programs.
\end{abstract}


\end{frontmatter}

\section{Introduction}\label{sec1}

In observational studies the assignment of units to treatments often
depends on latent variables that are related to the response variable
even when conditioning on known covariates. Consequently, a direct
comparison of the response distributions (given the model covariates) or
moments of these distributions between treatment groups could\vadjust{\eject} be biased
and result in wrong conclusions. An important example studied in this
article is the question of whether private schools offer better quality
of education than public schools. This question has important impact on
educational policy and public finance [\citet{Han02}]. It is known
that pupils enrolling to the two types of schools differ in their family
background and other characteristics related to their scholastic
achievements, such that a raw comparison of the scores of pupils
attending the two types of schools can be misleading. In an attempt to
deal with this question, we use data collected in the year 2000 by OECD
for the \textit{Programme for International Student Assessment} (PISA).
The purpose of this program is to study and compare the proficiency of
pupils aged 15 from over more than 30 countries in mathematics, science
and reading. In this article we focus on scores in mathematics in
Ireland and estimate the difference in the average score between the two
types of schools by adjusting for the quality of pupils enrolling to
them and for the effects of known covariates.

We start by applying several existing methods for observational studies
to the data, which are described in Section \ref{sec3}, and find, similarly to
Vanderberghe and Robin (\citeyear{VanRob04}), that some of these methods produce
estimates of different magnitude and sign. We attempt to resolve this
conflict by developing and applying a new method for inference from
observational data, which extends recent methodology for analyzing
sample survey data. The method derives the \textit{sample distribution}
of the observed response under a given treatment (score in mathematics
in a given type of school in our application) as a function of the
distribution that would be obtained under a strongly ignorable
assignment of subjects to treatments (assumptions $\operatorname{SI}(a)$, $\operatorname{SI}(b)$ in Section \ref{sec3}), and the assignment probability,
which is allowed to depend on the response value. The use of this
approach is established by showing that the sample distribution is
identifiable under some general conditions. The goodness of fit of the
sample distribution can be tested by standard test statistics since it
refers to the observed data.

By fitting the sample distribution to the observed data, we can estimate
the distribution under strongly ignorable assignment to treatments, and
the assignment probabilities, which are then used for estimating population means or contrasts between them. Our approach permits also
testing some of the assumptions underlying other methods for analyzing
observational data, thus enabling us to understand better why different
methods yield different answers in our application.

Section \ref{sec2} describes the PISA data and defines the problem underlying
this study more formally. Section \ref{sec3} overviews some of the existing
methods for observational studies and shows the results obtained when
applying them to the PISA data for Ireland. We also consider
``probability weighted'' versions of the estimators, which account for
unequal sample selection probabilities that are possibly related to the
response and may thus bias the inference. Computation of these
estimators yields very similar estimates\vadjust{\eject} to the estimates obtained under
the standard methods. Section \ref{sec4} presents the proposed approach and shows
the results obtained when applying it to the PISA data. The main outcome
of this analysis is that after controlling for the effect of the
enrollment process, the public schools actually outperform the private
schools in the average math score, suggesting better quality of
education. Here also we extend the method to account for unequal
selection probabilities and obtain similar estimates. Section \ref{sec5}
overviews the main theoretical properties of the new approach. The
technical derivations are presented in Supplements C and D of the
supplementary material [\citet{PfeLan}]. We conclude
with a brief summary in Section~\ref{sec6}.

\section{Data used for application and formulation of the problem}\label{sec2}

\subsection{Sampling design and response values}\label{sec2.1}

In order to compare the private and public schools, we use data
collected in \textit{Ireland} in the year 2000 by OECD for PISA.

\subsection*{Sampling design}

PISA uses in most countries a stratified two-stage sampling design. The
strata are defined by the size of the school, type of school and gender
composition. In each stratum, a probability proportional to size (PPS)
sample of schools was selected with the size defined by the number of
15-year-old pupils enrolled in the school. A minimum of 150 schools has
been selected in each country, or all the schools if there are less. In
the second stage an equal probability sample of 35 pupils from the
corresponding age group was drawn from each of the sampled schools (or
all the pupils in schools with less than 35 pupils aged 15). By this
sampling design, pupils included in the sample do not have equal
selection probabilities and each pupil is assigned therefore a sampling
weight. The weight is the reciprocal of the pupil's sample inclusion
probability, adjusted for nonparticipation of schools and nonresponse of
pupils.

PISA distinguishes between two types of private schools:
private-dependent schools where the government contributes 50\% or more
to the school core funding and private-independent schools with less
than 50\% government funding. The sample from Ireland consists of 54
public schools, 79 private-dependent schools and only 4
private-independent schools and, hence, in this paper we do not
distinguish between the two types of schools and refer to them simply as
private schools. For more information on the PISA sampling design and
weighting, see \citet{autokey23}.

\subsection*{Computation of response values}

The pupils' proficiencies (scores in mathematics in our case) are not
observed directly in the PISA study and are viewed as missing data,
which are imputed from the item responses $d_{j} =
(d_{1j},d_{2j},\ldots,d_{mj})$, where $d_{ij} = 1$ if pupil $j$ answers
correctly question $i$ of the examination and $d_{ij} = 0$ otherwise, $i
= 1,\ldots,m$. PISA uses two approaches for imputing the scores: a maximum\vadjust{\eject}
likelihood approach and a multiple imputation approach. In this paper we
used the imputed values obtained under the second approach. See
Appendix \ref{appA} for the imputation model. The PISA database contains five sets of
imputed values. We standardized the imputed scores in each set by
dividing them by their empirical standard deviation and then defined the
response value to be the average of the five standardized values. After
standardization and averaging, the range of the response values is
approximately from 1 to 10. We compared the use of the average values to
the results obtained when analyzing each of the five sets of
standardized values separately and then combining the results using
multiple imputation theory and obtained very similar results in all the
analyses performed. Consequently, in this paper we restrict to the
average response since it is convenient to have a single working model
when simulating new observations, which is needed for the
goodness-of-fit tests discussed later.

\subsection{Formulation of the problem}\label{sec2.2}

The formulation of the problem for the PISA data follows what is known
in the literature as the \textit{counterfactual approach}. By this
approach, every unit in the population is potentially exposed to every
treatment. See, for example, \citet{Rub74}, \citet{RosRub83},
\citet{SmiSug88} and \citet{Ros02}.

Let $U$ define the population of 15-year-old pupils in Ireland. Every
pupil $i \in U$ has two \textit{potential} responses: $Y_{1i}$---the
proficiency score if the student attends a private school, and $Y_{0i}
$---the proficiency score if the student attends a public school. Let $\mathrm{x}$
denote a set of $k$ known covariates (background characteristics) that
affect the responses, with values $\mathrm{x}_i$ for pupil $i$. The (potential)
population mean score in private schools (hereafter the treatment group)
is defined as $\mu_{1} = \frac{1}{N}\sum_{i = 1}^{N}
E_{p}[Y_{1i}|\mathrm{x}_i]$, where $N$ is the population size and the
expectation $E_{p}( \cdot)$ is with respect to the population model
holding for the responses. The population mean score in public schools
(hereafter the control group) is defined accordingly as $\mu_{0} =
\frac{1}{N}\sum_{i = 1}^{N} E_{p}[Y_{0i}|\mathrm{x}_i]$. In many observational
studies, contrasts between the parameters $\mu_{1}$ and $\mu_{0}$ are of
primary interest. In this paper we focus on estimating the difference
between the mean score in private and public schools, defined as
\begin{equation}\label{eq2.1}
\tau = \mu_{1} - \mu_{0} = \frac{1}{N}\sum_{i = 1}^{N}
E_{p}[Y_{1i}|\mathrm{x}_i] - \frac{1}{N}\sum_{i = 1}^{N} E_{p}[Y_{0i}|\mathrm{x}_i].
\end{equation}
The contrast $\tau$ is known in the literature as the \textit{average
treatment effect} (ATE).

In practice, every unit in the population is only exposed to one
treatment. Also, it is rarely the case that all the population units
participate in the study. The observed data refer therefore to a sample
$S$ of size $n$, which in our application is divided into the two
subsamples $S_{1}$ and $S_{0}$, where $S_{1}$ $(S_{0})$ is the subsample of
pupils attending private (public) schools. For every pupil $i \in S$ we
observe therefore $y_{1i}$ if $i \in S_{1}$ or $y_{0i}$ if $i \in
S_{0}$.\vadjust{\eject}

Denote by $\pi_{i}$ the probability that pupil $i$ is included in the
sample $S$ and by $\tilde{p}_{t i}$ the probability that
\textit{sampled} pupil $i$ is enrolled in school of type $t$. The sample
inclusion probabilities $\pi_{i} $ (or the sampling weights $w_{i} =
1/\pi_{i}$, with possible adjustments for nonresponse or calibration)
are typically known for the sampled units, as is the case in the PISA
survey, but the treatment assignment probabilities, $\tilde{p}_{t i}$, are usually unknown and may depend on latent variables that
are related to the response, $Y_{t i}$. As is well known and
illustrated later, if the effect of these latent variables on the
response is not accounted for by the observed covariates, the resulting
estimators of the population parameters can be highly biased.

\begin{remark}\label{rem1}
The sample inclusion probabilities may
likewise be related to the response values and thus bias the inference
if not accounted for adequately. This is known in the survey sampling
literature as \textit{informative sampling}. \citet{SmiSug88}
define conditions on the sampling design and the treatment assignment
process that warrant ignoring them in the inference process. As shown in
subsequent sections, there is no evidence for informative sampling with
the kind of models and inference methods applied to the PISA data from
Ireland.
\end{remark}

\section{Existing methods, application to PISA data}\label{sec3}

In what follows we focus on the estimation of the ATE defined by (\ref{eq2.1}),
assuming that the sample selection probabilities are not related to the
response variable and the covariates, and hence that there are no
sampling effects. This is the common assumption in the literature even
though seldom stated explicitly. After describing several methods in
common use and applying them to the PISA data, we show the results
obtained when extending the methods to the case where the sample is
selected with known unequal probabilities that might be related to the
response and/or the covariates and compare the results with the results obtained when ignoring
the sample selection. Let~$T$ define the indicator of the treatment
group ($T = 1$ for private schools, $T = 0$ for public schools). \citet{RosRub83} establish two conditions that warrant ignoring the
treatment assignment in the inference process
when conditioning on x:
\begin{itemize}
\item[$\bullet$]  $\operatorname{SI}(a)$: the assignment $T$ and the response
values $(Y_{1},Y_{0})$ are independent given the covariates,
$\mathrm{x}$, for every unit (pupil),

\item[$\bullet$]  $\operatorname{SI}(b)$: $0 < \Pr(T = 1|\mathrm{x}) < 1$ for every
possible $\mathrm{x}$.

\end{itemize}
Conditions $\operatorname{SI}(a)$ and $\operatorname{SI}(b)$ define a \textit{strongly
ignorable assignment process} given the covariates. When the assignment
is strongly ignorable, it permits the application of a number of simple
estimation techniques, which we review in Section \ref{sec3.1}. In Sections \ref{sec3.2}
and \ref{sec3.3} we consider the latent variables method (LV) and the use of
instrumental variables (IV), which do not assume strong ignorability
assumptions.

\subsection{Methods for strongly ignorable treatment assignments}\label{sec3.1}

\mbox{}

\textit{Regression estimator}. Suppose that the true
relationship between $Y$ and $\mathrm{x}$ in the population has the
general form $Y_{t} = r_{t}(\mathrm{x}) + u_{t}$, $E_{p}(u_{t}|\mathrm{x}) = 0$ for some
functions $r_{t}(\mathrm{x})$, $t = 0,1$, where $E_{p}( \cdot)$ is the expectation
under a strongly ignorable assignment. Then, the ATE is $(\bar{r}_{1} -
\bar{r}_{0})$, where $\bar{r}_{t} = \frac{1}{N}\sum_{i = 1}^{N}
r_{t}(\mathrm{x}_i)$. When the expectations $r_{t}(\mathrm{x})$ are linear, $r_{t}(\mathrm{x}) =
\mathrm{x}'\beta_{t}$, then under the assumptions $\operatorname{SI}(a)$, $\operatorname{SI}(b)$ the
regression coefficients can be estimated by ordinary least squares (OLS)
and the ATE estimator takes the form
\begin{equation}\label{eq3.1}
\hat{\tau} _{\mathrm{OLS}} = \bar{\mathrm{x}}'(\hat{\beta}
_{1,\mathrm{OLS}} - \hat{\beta} _{0,\mathrm{OLS}}),
\end{equation}
where $\mathrm{\bar{x}'} =
(\mathrm{\bar{x}}_{1},\ldots,\bar{\mathrm{x}}_{K}) = \sum_{i =
1}^{n} \mathrm{x}_{i}/n$ and $\hat{\beta} _{t,\mathrm{OLS}}$ is the OLS
estimator in subsample $S_{t}$.

\textit{Matching estimator}. Another procedure in common use
is to match the units from the treatment and control groups based on the
covariates and then compare the responses. Matching procedures are
widely discussed in the literature; see, for example, \citet{Ros02}.
They do not require specifying the form of the functions $r_{t}(\mathrm{x})$.
\citet{AbaImb06} consider the following matching estimate with
replacement. Denote by $J_{iM}(t)$ the indices of the $M$ closest
matches in $S_{t}$ for unit $i \in S_{1 - t}$, $t = 0,1$. Define for unit $i
\in S_{1 - t}, \hat{y}_{(1 - t), i} = y_{(1 - t),
i}$ and $\hat{y}_{t i} = \frac{1}{M}\sum_{j \in J_{iM}(t)}
y_{t j}$. Estimate
\begin{equation}\label{eq3.2}
\hat{\tau} _{\mathrm{M}} = \frac{1}{n}\sum_{i = 1}^{n}
(\hat{y}_{1 i} - \hat{y}_{0 i}).
\end{equation}

Other methods use probability weighting with the weights defined by the
inverse of the ``propensity score,'' $e(\mathrm{x}) = \Pr(T = 1|\mathrm{x})$. \citet{RosRub83} show that the conditions
$\operatorname{SI}(a)$, $\operatorname{SI}(b)$ for
strong ignorability imply the same conditions when $\mathrm{x}$ is replaced
by $e(\mathrm{x})$, thus validating the use of propensity scores for ATE
estimation. In practice, the propensity scores are unknown and are
estimated by fitting logistic or probit models, or by use of
nonparametric techniques [McCaffrey, Ridgeway and Morral (\citeyear{McCRidMor04})]. Below
we describe two ATE estimators that use the
estimated propensity scores, $\hat{e}({\mathrm{x}}_{i})$, for weighting.

\textit{Brewer--Hajek} (\textit{B--H}) \textit{estimator}.
This estimator resembles the familiar Brewer--Hajek [\citet{Bre63}; Hajek
(\citeyear{Haj71})] estimator in survey sampling. Let $T_{i} = 1 (0)$ if unit $i \in
S_{1}$ $(i \in S_{0})$ and define $Y_{i} = T_{i}Y_{1 i} + (1 -
T_{i})Y_{0i}$.
The B--H estimator is
\begin{equation}\label{eq3.3}
\hspace*{27pt} \hat{\tau} _{\mathrm{B \mbox{--} H}} = \Biggl( \sum_{i = 1}^{n}
\frac{T_{i}}{\hat{e}(\mathrm{x}_i)} \Biggr)^{ - 1}\sum_{i = 1}^{n}
\frac{T_{i}Y_{i}}{\hat{e}(\mathrm{x}_i)} - \Biggl( \sum_{i = 1}^{n} \frac{1 -
T_{i}}{1 - \hat{e}(\mathrm{x}_i)} \Biggr)^{ - 1}\sum_{i = 1}^{n} \frac{(1 -
T_{i})Y_{i}}{1 - \hat{e}(\mathrm{x}_i)}.
\end{equation}

\textit{Doubly-robust} (\textit{DR}) \textit{estimator}. If
the population expectation $E_{p}(Y_{t}|\mathrm{x})$ can be modeled by some
function $r_{t}(\mathrm{x})$, then by $\operatorname{SI}(a)$, $r_{t}(\mathrm{x})$ is also the sample
expectation and the ATE can be estimated as
\begin{eqnarray} \label{eq3.4}
\hat{\tau} _{\mathrm{DR}} &=& \frac{1}{n}\sum_{i = 1}^{n}
\frac{T_{i}Y_{i} - [T_{i} -
\hat{e}(\mathrm{x}_i)]\hat{r}_{1}(\mathrm{x}_i)}{\hat{e}(\mathrm{x}_i)}\nonumber\\[-8pt]\\[-8pt]
&&{} - \frac{1}{n}\sum_{i= 1}^{n} \frac{(1 - T_{i})Y_{i} + [T_{i} -
\hat{e}(\mathrm{x}_i)]\hat{r}_{0}(\mathrm{x}_i)}{1 - \hat{e}(\mathrm{x}_i)}.\nonumber
\end{eqnarray}
The estimator (\ref{eq3.4}) has the ``double-robustness'' property of being
consistent even if only the model assumed for the propensity scores or
the population expectations are correctly specified [Lunceford and
Davidian (\citeyear{LunDav04})]. \citet{QinZha07} consider another estimator that
has a somewhat stronger robustness property.

\subsection{Latent variable models}\label{sec3.2}

This method specifies the joint distribution of the outcome and the
treatment selection by use of latent variable (LV) models. The model
assumes the following:

\begin{itemize}
\item  $\operatorname{LV}(a)$---a structural equation for the
population outcomes of the form, $Y_{t} = r_{t}(\mathrm{x}) + u_{t};E_{p}(u_{t}|\mathrm{x})
= 0, t = 0,1$, and

\item  $\operatorname{LV}(b)$---a latent variable $u_{T}$ and an
assignment rule $T$ satisfying $T = 1 [l(\mathrm{v};\alpha) + u_{T} >
0]$, where $1 [ \cdot ]$ is the indicator function and
$l(\mathrm{v};\alpha)$ is a~given function of known
covariates $\mathrm{v}$, governed by a vector parameter  $\alpha$.
\end{itemize}

The
covariates in $\mathrm{v}$ may include some of the covariates
in $\mathrm{x}$, but it is generally recommended that $\mathrm{v}
\ne\mathrm{x}$ to avoid colinearity problems in the estimation process;
see below. The random variables $(u_{1},u_{0},u_{T})$ are dependent.
Under these assumptions, $E_{S_{t}}(Y_{t}|\mathrm{x}) =E(Y_{t}|\mathrm{x},T = t)=
r_{t}(\mathrm{x})+ E(u_{t}|\mathrm{x},T = t)\ne r_{t}(\mathrm{x}) = E_{p}(Y_{t}|\mathrm{x})$,
since $E(u_{t}|\mathrm{x}, T = t) \ne 0$. However, assuming that $r_{t}(\mathrm{x}) =
\mathrm{x}'\beta_{t}, (u_{1},u_{0},u_{T})$ are jointly normal
and $l(\mathrm{v};\alpha) = \mathrm{v}'\alpha, \beta_{t}$ can be
estimated by the two-stage Heckman's method [\citet{Mad83}], yielding

\begin{equation}\label{eq3.5}
\hat{\tau} _{\mathrm{LV}} = \bar{\mathrm{x}}'(\hat{\beta}
_{1,\mathrm{LV}} - \hat{\beta} _{0,\mathrm{LV}}),
\end{equation}
where $\hat{\beta} _{t,\mathrm{LV}}$ is the LV estimator of $\beta_{t},t
= 0,1$. \citeauthor{HecVyt06} [(\citeyear{HecVyt06}), Ch.~70] refer to the latter model as the
Generalized Roy Model and discuss semi-parametric econometric models,
which relax some of the assumptions of this model.

\subsection{Instrumental variables models}\label{sec3.3}

Let $Y_{t} = \mathrm{x}'\beta_{t} + u_{t};E_{p}(u_{t}|\mathrm{x}) = 0,t =
0,1$. Then for unit $i \in S$,
\begin{equation}\label{eq3.6}
Y_{i} = T_{i}\mathrm{x}'_{i}\beta_{1} + (1 -
T_{i})\mathrm{x}'_{i}\beta_{0} + u_{i} = \mathrm{\tilde{x}'}_{i}\theta +
u_{i},
\end{equation}
where $Y_{i} = T_{i}Y_{1i} + (1 - T_{i})Y_{0i}$, $\mathrm{\tilde{x}'}_{i} =
(T_{i}\mathrm{x}'_{i},(1 - T_{i})\mathrm{x}'_{i})$, $\theta =
(\beta_{1}^{\prime} ,\beta_{0}^{\prime} )'$ and $u_{i} = T_{i}u_{1i} +
(1 - T_{i})u_{0i}= u_{0i} + T_{i}(u_{1i} - u_{0i})$. In observational
studies $u_{1}$ and $u_{0}$ are\vadjust{\eject} correlated with $T$ and, hence,
$\theta$ cannot be estimated consistently from (\ref{eq3.6}) without additional
assumptions. Below we define a set of plausible assumptions warranting
that the ATE is estimated consistently. See \citet{Woo02} for
discussion of this and alternative sets of assumptions. Assume the
availability of an instrument $h$ satisfying the following:
\begin{itemize}
\item IV(a)---$E_{p}(Y_{t}|\mathrm{x},h) =
E_{p}(Y_{t}|\mathrm{x})$ (the population expectation under strongly ignorable assignment does not depend on
the instrument, given the covariates);

\item  IV(b)---$E_{p}[T(u_{1} - u_{0})|\mathrm{x},h] =
0$ (the assignment and the counterfactual gain in the error terms are
uncorrelated given the covariates and the instrument);

\item IV(c)---$\Pr(T = 1|\mathrm{x},h)=
g(\mathrm{x},h) \ne\Pr(T = 1|\mathrm{x}) $ (the assignment probabilities
depend on the instrument and possibly on $\mathrm{x}$).
\end{itemize}

Multiplying both sides of (\ref{eq3.6}) by the column vector $\mathrm{z}_{i} =
(g_{i}\mathrm{x}'_{i},(1 - g_{i})\mathrm{x}'_{i})'$, where $g_{i} =
g(\mathrm{x}_{i},h_{i})$ and taking expectations yields $E_{p}(\mathrm{z}_{i}
Y_{i}|\mathrm{x}_{i}, h_{i}) = E_{p}(\mathrm{z}_{i}
\mathrm{\tilde{x}'}_{i}|\mathrm{x}_{i}, h_{i})\theta$, since under the
model and IV(b), $E_{p}(\mathrm{z}_{i}u_{i}|\mathrm{x}_{i},h_{i}) = 0$. The IV
estimator of $\theta$ computed from all the observations is $\hat{\theta}
_{\mathrm{IV}} =\break (\sum_{i = 1}^{n} \hat{\mathrm{z}}_{i}\tilde{\mathrm{x}}'_{i})^{ -
1}\sum_{i = 1}^{n} \hat{\mathrm{z}}_{i}Y_{i}$, where $\hat{\mathrm{z}}_{i} =
(\hat{g}_{i}\mathrm{x}'_{i},(1 - \hat{g}_{i})\mathrm{x}'_{i})'$. The
estimator $\hat{g}_{i} = \hat{g}(\mathrm{x}_{i},h_{i})$ is commonly
obtained by fitting probit or logit models. The ATE estimator is
\begin{equation}\label{eq3.7}
\hat{\tau} _{\mathrm{IV}} = \bar{\mathrm{x}}'(\hat{\beta} _{1,\mathrm{IV}}
- \hat{\beta} _{0,\mathrm{IV}}),
\end{equation}
with $\hat{\beta} _{t,\mathrm{IV}}$ defined by $\hat{\theta}
_{\mathrm{IV}},t = 0,1$.

\begin{remark}\label{rem2}
Condition IV(c) is testable from the data, but
conditions IV(a) and IV(b) relate to unobservable quantities and cannot
in general be tested directly. \citet{ImbAng94} show that for a
binary instrument, if condition IV(b) is not satisfied, then under a
weaker \textit{monotonicity condition}, $\hat{\tau} _{\mathrm{IV}}$ estimates the
treatment effect for a subpopulation consisting of units for which the
treatment status would be altered by the instrument. This treatment
effect is called \textit{local average treatment effect} (LATE).
\end{remark}

\subsection{Application of the methods to PISA data for Ireland}\label{sec3.4}

We applied the methods reviewed so far to the PISA data for Ireland
described in Section \ref{sec2}. The sample consists of 1,244 pupils from private
schools and 694 pupils from public schools. Six covariates were found to
be significant in at least one of the models described in Section \ref{sec4}:
gender (GEN; 1 for girls, 0 for boys), mother's education (ME; 1 for
high education, 0 otherwise), family socio-economic index (SEI), index
of home educational resources (HER), average socio-economic index of
the pupil's schoolmates [SES; proposed by \citet{VanRob04} to account for potential peer effects],
and school location (S.loc; 1 if the school is located in an urban area, 0 otherwise). The
continuous covariates have\vadjust{\eject} been standardized. To warrant fair
comparability between the various methods, we included for the first
four methods [equations (\ref{sec3.1})--(\ref{eq3.4})] all the six covariates in both the
regressions and the models used for computing the propensity scores. For
the LV and IV methods we included all the covariates except for S.loc in
the regressions and all the covariates including S.loc in the school
selection models (see Remark 3). \citet{VanRob04}
considered additional covariates, but these were not found to be
significant in our analysis.

\begin{table}
\caption{Estimates of ATE and standard errors under existing methods}
\label{tab1}
\begin{tabular}{@{}lccccccc@{}}
\hline
\textbf{Method} & $\bolds{\hat{\tau} _{\mathrm{D}}}$ & $\bolds{\hat{\tau} _{\mathrm{OLS}}}$ & $\bolds{\hat{\tau} _{\mathrm{M}}}$ &
$\bolds{\hat{\tau} _{\mathrm{B \mbox{\textbf{--}} H}}}$ & $\bolds{\hat{\tau} _{\mathrm{DR}}}$ & $\bolds{\hat{\tau} _{\mathrm{LV}}}$ & $\bolds{\hat{\tau} _{\mathrm{IV}}}$\\
\hline
Estimate & 0.36 & 0.12 & 0.21 & 0.16 & 0.17 & $-$0.49 & $-$0.61\\
Std. error & 0.05 & 0.05 & 0.05 & 0.05 & 0.05 & \hphantom{$-$}0.19 & \hphantom{$-$}0.24\\
\hline
\end{tabular}
\end{table}

\begin{remark}\label{rem3}
The variable \textit{school location} was used by
\citet{VanRob04} as an instrumental variable. The authors
show that it has a significant effect on the probability of attending
private schools, thus satisfying the condition IV(c) in Section \ref{sec3.3}.
However, the approaches considered in the literature for observational
studies do not permit testing that the school location is exogenous to
the pupil's proficiency given the other covariates, as required by
condition IV(a), because this condition refers to the population models
of the unobservable potential responses. The authors argue that this
requirement is plausible, using similar arguments to \citet{Hox00}. See
Section \ref{sec4.6} for how we can test this condition under the approach
proposed in Section \ref{sec4}.
\end{remark}

Table \ref{tab1} presents the ATE estimates and their standard errors. The first
estimate, $\hat{\tau} _{\mathrm{D}} = \bar{y}_{1} - \bar{y}_{0}$, is the
crude difference between the simple sample means in the two types of
schools. The matching estimator is computed based on $M = 4$ matches. We
considered several matching estimates as obtained under different
metrics for finding the matches, with and without adjustments for
imperfect matching, and obtained very close results in all the cases.
For the instrumental variables method we used the school location as the
instrument.

The estimates $\hat{\tau} _{\mathrm{M}},\hat{\tau} _{\mathrm{LV}}$ and
$\hat{\tau} _{\mathrm{IV}}$ were computed by using the functions
\textit{nnmatch}, \textit{treatreg} and \textit{ivreg} of the Stata
software [\citet{autokey33}]. The remaining estimates were programmed
using the R software [\citet{autokey25}]. Estimation of the
standard errors of the matching estimators and the LV and IV estimators
is incorporated in the Stata functions. See \citet{AbaImb06}
and \citet{Woo02} for details. Estimation of the standard errors of
the Brewer--Hajek estimator and the doubly robust estimator\vadjust{\eject} is developed
by Lunceford and Davidian (\citeyear{LunDav04}).
The estimated standard errors account for the error distributions of the responses under the respective models.

The first notable outcome in Table \ref{tab1} is that the difference $\hat{\tau}
_{\mathrm{D}}$ between the simple sample means in the two types of
schools is positive, which we anticipated because the more able pupils
tend to enroll in private schools. The next four methods from left,
which assume strongly ignorable assignment given the covariates,
likewise produce small positive ATE estimates. By contrast, the IV and
LV methods, which account for treatment assignment effects not explained
by the covariates, produce negative estimates, with much larger absolute
values, suggesting that the public schools actually perform better after
accounting for the school selection effects. A~similar outcome is obtained
under the approach proposed in Section \ref{sec4}. The use of this approach
explains also why the LV and IV methods are more appropriate for this
data.\looseness=-1

\begin{remark}\label{rem4}
\citet{VanRob04} computed what is known
in the econometric literature as ``the average treatment effect for the
treated (ATT),'' using the same data and some of the methods reviewed
before, and obtained similar results to the results in Table \ref{tab1}. \citet{DroAvr10} computed the ATT for reading scores using PISA data for
all the countries by applying several variants of propensity scores
matching. The ATT estimates for Ireland in this study are positive, same
as the ATE estimates for the scores in Mathematics based on propensity
scores presented in Table \ref{tab1} ($\hat{\tau} _{\mathrm{B -
H}}$ and $\hat{\tau} _{\mathrm{DR}}$).
\end{remark}

\subsection{Probability weighted estimators for PISA data}\label{sec3.5}

So far we ignored the sample selection process when computing the
estimates in Section \ref{sec3.4}. The question arising is whether this is
justified in the present study. We emphasize again that if the
distribution of the response in the sample is affected by the sample
selection scheme, the sampling is informative and failing to account for
the sampling effects may bias the inference. In fact, even if only the
distribution of the covariates in the sample is different from their
population distribution, some of the ATE estimators may already be
biased. \citet{PfeSve09} review several existing
approaches to account for possible sampling effects in the inference
process. In this study we applied what is known as probability
weighting, which basically consists of inflating each sample observation
proportionally to its sampling weight. The idea of probability weighting
is to obtain estimators that are consistent under the randomization
(repeated sampling) distribution for the corresponding ``census
estimates'' that would be computed if all the population values had been
observed. The census estimates are free of sampling effects.

We computed the probability weighted estimators (PWE) for all the
methods considered so far. See Supplement A in the
supplementary
material [\citet{PfeLan}] for the\vadjust{\eject} derivation of these
estimators. As a first step we computed the unweighted and probability
weighted estimators (in parenthesis) of the population means of the
covariates and obtained the following results (the covariates are
defined in Section \ref{sec3.4}): GEN: 0.53 (0.52), ME: 0.61 (0.61), SEI: 0.00
($-$0.016), SES: 0.00 ($-$0.016), HER: 0.00 (0.002), S.loc: 0.40 (0.39).
As can be seen, the two sets of estimators are very close.

\begin{table}[t]
\caption{Unweighted and weighted estimators of schools mean score and ATE}
\label{tab2}
\vspace*{-2pt}
\begin{tabular*}{\tablewidth}{@{\extracolsep{\fill}}lcccccc@{}}
\hline
& \multicolumn{2}{c}{\textbf{Private schools}} & \multicolumn{2}{c}{\textbf{Public schools}} & \multicolumn{2}{c@{}}{\textbf{ATE}}\\[-5pt]
&\multicolumn{2}{c}{\hrulefill}&\multicolumn{2}{c}{\hrulefill}&\multicolumn{2}{c@{}}{\hrulefill}\\
\textbf{Method}  & \textbf{UNWEI} & \textbf{WEI} & \textbf{UNWEI} & \textbf{WEI} & \textbf{UNWEI} & \textbf{WEI}\\
 \hline
Simple difference & 6.28 & 6.29 & 5.92 & 5.92 & \hphantom{$-$}0.36 & \hphantom{$-$}0.37\\
Regression & 6.21 & 6.26 & 6.09 & 6.12 & \hphantom{$-$}0.12 & \hphantom{$-$}0.14\\
Matching & 6.25 & 6.26 & 6.04 & 6.03 & \hphantom{$-$}0.21 & \hphantom{$-$}0.23\\
Brewer--Hajek (B--H) & 6.24 & 6.26 & 6.08 & 6.06 & \hphantom{$-$}0.16 & \hphantom{$-$}0.20\\
Doubly robust (DR) & 6.23 & 6.25 & 6.06 & 6.07 & \hphantom{$-$}0.17 & \hphantom{$-$}0.18\\
Instrumental variable & 6.00 & 6.02 & 6.61 & 6.52 & $-$0.61 & $-$0.50\\
Latent variable & 6.00 & 6.02 & 6.49 & 6.41 & $-$0.49 & $-$0.39\\
\hline
\end{tabular*}
\vspace*{-2pt}
\end{table}

Table \ref{tab2} shows for each of the methods the unweighted (UNWEI) and
probability weighted (WEI) estimators of the mean score in the private
and public schools, and the corresponding ATE estimator. The results in
Table~\ref{tab2} indicate that the PWE of the mean score as obtained under the
various methods are very similar to the corresponding unweighted
estimators. This is definitely true for the private schools, but even
for the public schools the largest difference between the weighted and
unweighted estimate is less than 2\%. The very small differences
between the weighted and unweighted estimates in each type of school
translate into somewhat larger differences in the estimates of the ATE,
but not to an extent that affects the inference. Notice in this regard
that when computing the conventional 95\% confidence intervals for the
true ATE based on the unweighted ATE estimates, all the intervals
contain the corresponding weighted estimates. In fact, this would be
the case even for confidence intervals with confidence level as low as
68\%. Our general conclusion from Table \ref{tab2} is therefore that the sampling
process can be ignored when analyzing the PISA data from Ireland by use
of the methods considered so far.

\section{An alternative approach for observational studies}\label{sec4}

In this section we propose an alternative approach for ATE estimation,
which, as illustrated in Section \ref{sec4.6}, allows also testing the
appropriateness of candidate instrumental variables or the use of
propensity scores under the assumed model. The approach resembles the LV
approach in the sense that it assumes a~population model and a model for
the treatment selection and applies a~combined likelihood resulting from
the two models, but all the\vadjust{\eject} subsequent developments are very different.
As with the IV and LV methods, the use of this approach does not require
strong ignorability assumptions. In what follows we describe the method
and apply it to the PISA data assuming noninformative sampling, but
later we also consider probability weighted estimation. As before, we
consider the case of two groups, $t = 0,1$.

\subsection{The sample distribution}\label{sec4.1}

Denote by $f_{p}(y_{t i}|\mathrm{x}_{i})$ the
\textit{population pdf} for units in treatment group $t$ under
a strongly ignorable assignment process. We allow the assignment process
to depend on known covariates $\mathrm{v}$, some or all of which may be
included in $\mathrm{x}$. Denoting $\mathrm{z} = \mathrm{x} \cup\mathrm{v}$, we
assume $f_{p}(y_{t i}|\mathrm{z}_{i}) = f_{p}(y_{t
i}|\mathrm{x}_{i})$ and $\Pr(T_{i} = t|y_{t
i},\mathrm{z}_{i},i \in S)= \Pr(T_{i} = t|y_{t
i},\mathrm{v}_{i},i \in S)$. The \textit{sample} \textit{pdf} for unit
$i$ exposed to treatment $t$, given the covariates $\mathrm{z}_{i}$, is obtained by
Bayes theorem as
\begin{equation}\label{eq4.1}
f_{S_{t}}(y_{t i}|\mathrm{z}_{i}) =
f(y_{t i}|\mathrm{z}_{i}; T_{i} = t) = \frac{\Pr (T_{i} =
t|y_{t i},\mathrm{v}_{i},i \in S)f_{p}(y_{t
i}|\mathrm{x}_{i})}{\Pr (T_{i} = t|\mathrm{z}_{i},i \in S)},
\end{equation}
where $\Pr(T_{i} = t|\mathrm{z}_{i},i \in S) = \int\Pr(T_{i} =
t|y_{t i},\mathrm{v}_{i},i \in S)f_{p}(y_{t
i}|\mathrm{x}_{i})\,dy_{ti}.$

\begin{remark}\label{rem5}
It follows from (\ref{eq4.1}) that the sample \textit{pdf} is
generally different from the population \textit{pdf}, unless $\Pr(T_{i}
= t|y_{t i}, \mathrm{v}_{i},i \in S)= \Pr(T_{i} = t|
\mathrm{z}_{i},i \in S)$, in which case the assignment to treatments can
be ignored for inference when conditioning on $\mathrm{z}$.
\end{remark}

\begin{remark}\label{rem6}
The probabilities $\Pr(T_{i} = t| \mathrm{z}_{i},i
\in S)$ are \textit{propensity scores}.
\end{remark}

The sample \textit{pdf} defined by (\ref{eq4.1}) was shown in recent years to
provide a valuable modeling approach for inference from complex sample
surveys; see \citet{PfeSve09} for review of studies that utilize
the sample \textit{pdf} for inference on generalized linear models,
testing of distribution functions and prediction of finite population
and small area means. The obvious distinction between survey sampling
and observational studies is that in survey sampling the sample
inclusion probabilities $\pi_{i} = \Pr(i \in S)$ are usually known,
which enables estimating the probabilities $\Pr(i \in S|y_{i} ,
\mathrm{v}_{i})$ and testing the
informativeness of the sampling process [Pfeffermann and Sverchkov
(\citeyear{PfeSve03}, \citeyear{PfeSve09})]. This is generally not the
case in observational studies, requiring therefore modeling the
parametric form of the probabilities $p_{t i} = \Pr(T_{i} = t|y_{t i} ,
\mathrm{v}_{i},i \in S)$ in (\ref{eq4.1}). As discussed below, modeling the
sample \textit{pdf} (\ref{eq4.1}) allows estimating the unknown parameters
governing the \textit{pdf} $f_{p}(y_{t i}|\mathrm{x}_{i})$ and the
probabilities $p_{t i}$, and using them for estimating the ATE.

\subsection{Estimating the parameters of the sample distribution}\label{sec4.2}

So far we suppressed for convenience in the notation the parameters
governing the sample \textit{pdf}. Adding the unknown parameters to the
notation and assuming that the inclusion\vadjust{\eject} in the sample and the
assignment to treatments are independent between units, and that the
responses are likewise independent, the sample likelihood for treatment
$t$, based on the sample $S_{t}$ of size $n_{t}$, takes the\break form
\begin{equation}\label{eq4.2}
L_{S_{t}}[\alpha_{t}, \theta_{t}; \{ y_{t i} ,
\mathrm{z}_{i}\} ] = \prod_{i = 1}^{n_{t}} \frac{\Pr (T_{i} =
t|y_{t i},\mathrm{v}_{i},i \in S;\alpha _{t})f_{p}(y_{t i}|\mathrm{x}_{i} ;
\theta _{t})}{\Pr (T_{i} = t|\mathrm{z}_{i},i
\in S;\alpha _{t},\theta _{t})}.
\end{equation}

Alternatively, the likelihood (\ref{eq4.2}) can be replaced by the joint
(``full'') likelihood of the treatment selection and the response
measurements defined as
\begin{eqnarray}\label{eq4.3}
&&L_{S}[\alpha_{t} , \theta_{t}; \{ y_{t i} ,i \in S_{t}; \mathrm{z}_{j}, j \in S\} ] \nonumber \\
&&\qquad = \prod_{i =
1}^{n_{t}} \Pr(T_{i} = t|y_{t i},\mathrm{v}_{i},i \in
S;\alpha_{t})f_{p}(y_{t  i}|\mathrm{x}_{i} ;
\theta_{t})\\
&&\qquad \quad {}\times\mathop{\prod_{j \in S }}_{j \notin
S_{t}} [1 - \Pr(T_{j} = t|\mathrm{z}_{j},j \in S;\alpha_{t},\theta_{t})].\nonumber
\end{eqnarray}
The likelihood (\ref{eq4.3}) has the advantage of comprising the unconditional
treatment assignment probabilities for units outside the sample $S_{t}$,
thus using more information for estimating the model parameters. The
full likelihood is often applied for handling informative nonresponse;
see, for example, Greenlees, Reece and Zieschang (\citeyear{GreReeZie82}),
\citet{RotRob97}, Gelman et al. [(\citeyear{Geletal04}), Chapter 7] and \citet{Lit04}.

Replacing the unknown model parameters by their maximum likelihood
estimates (\textit{mle}) yields the estimates
\begin{equation}\label{eq4.4}
\hat{f}_{p}(y_{t  i}|\mathrm{x}_{i}) = f_{p}(y_{t  i}|\mathrm{x}_{i} ; \hat{\theta} _{t});\qquad
\hat{p}_{t  i} = \Pr(T_{i} = t|y_{t  i},\mathrm{v}_{i},i \in S;\hat{\alpha}_{t}).
\end{equation}

\subsection{Estimation of population means}\label{sec4.3}

When the covariates $\mathrm{x}_{i}$ are known for every unit $i \in U$,
then by (\ref{eq4.4}), the population means, $\mu_{t} = \frac{1}{N}\sum_{i =
1}^{N} E_{p}(Y_{t  i}|\break\mathrm{x}_{i})$, $t = 0,1$ (and,
consequently, the ATE, $\tau = \mu_{1} - \mu_{0}$) can be estimated\break by
\begin{equation}\label{eq4.5}
\hat{\mu} _{t,p} = \frac{1}{N}\sum_{i = 1}^{N}
\hat{E}_{p}(Y_{t  i}|\mathrm{x}_{i};\theta_{t}) =
\frac{1}{N}\sum_{i = 1}^{N} E_{p}(Y_{t  i}|\mathrm{x}_{i} ;
\hat{\theta} _{t}).
\end{equation}
Note that if $E_{p}(Y_{t  i}|\mathrm{x}_{i} ; \theta_{t})$ is
linear, the computation of (\ref{eq4.5}) only requires knowledge of the
population means $\bar{X} = (\bar{X}_{1},\ldots,\bar{X}_{k})'$.

When the covariates are unknown for units outside the sample, or the
expectation is not linear, then as long as the sampling design is
noninformative with respect to the distribution of the covariates, one
can use the estimator,
\begin{equation}\label{eq4.6}
\hat{\mu} _{t,S} = \frac{1}{n}\sum_{i \in S}
\hat{E}_{p}(Y_{t  i}|\mathrm{x}_{i};\theta_{t}).
\end{equation}
Alternatively, one can use in this case the ``combined'' estimator,
\begin{equation}\label{eq4.7}
\hat{\mu} _{t,C} = \frac{\sum_{i = 1}^{n} T_{i}[Y_{t
i} - \hat{E}_{p}(Y_{t  i}|\mathrm{x}_{i} ; \theta
_{t})]/\hat{p}_{t  i}} {\sum_{i = 1}^{n}
(T_{i}/\hat{p}_{t  i})} + \hat{\mu} _{t,S}.
\end{equation}

\begin{remark}\label{rem7}
The estimator (\ref{eq4.7}) resembles the familiar GREG
estimator in survey sampling [S\"{a}rndal, Swensson and Wretman (\citeyear{SarSweWre92})],
and it looks similar to the ``doubly-robust'' estimator (\ref{eq3.4}). Notice,
however, that (\ref{eq4.7}) accounts for an informative assignment process as
reflected by the use of the probabilities $\hat{p}_{t  i} =
\operatorname{\hat{P}r}(T_{i} = t|y_{t  i} , \mathrm{v}_{i},i \in S)$ instead
of the propensity scores $\hat{e}_{ti} = \operatorname{\hat{P}r}(T_{i} =
t|\mathrm{z}_{i},i \in S)$. On the other hand, the estimator (\ref{eq4.7}) does
not posses a ``double robustness'' property, since the unknown model
parameters are estimated jointly from the likelihood (\ref{eq4.3}).
\end{remark}

The estimators (\ref{eq4.5}) and (\ref{eq4.6}) are functions of the
\textit{mle} $\hat{\theta} _{t}$, and, hence, their large sample variance
can be estimated by use of the inverse of the estimated information
matrix. Large sample properties of the combined estimator (\ref{eq4.7}) and a
consistent estimator of its variance can be derived by application of
$M$-estimation theory, see \citet{Lan} for details. The estimated
variances of all the three estimators account for the sample
distribution of the responses given the observed covariates.

\subsection{Application of new method to PISA data for Ireland}\label{sec4.4}

We assume a~normal distribution for the potential population responses
under strongly ignorable assignment and a logistic model for the
assignment probabilities:
\begin{eqnarray}\label{eq4.8}
f_{p}(y_{t  i}|\mathrm{x}_{i}) &=& N(\beta_{0t} +
\mathrm{x}'_{i}\beta_{t},\sigma_{t}^{2});\nonumber\\[-8pt]\\[-8pt]
\qquad \Pr(T_{i} = t|y_{t
i},\mathrm{v}_{i},i \in S) &=& \frac{\exp (\gamma _{0t} + \delta
_{t}y_{t  i} + \mathrm{v}'_{i}\gamma _{t})}{1 + \exp (\gamma
_{0t} + \delta _{t}y_{t  i} + \mathrm{v}'_{i}\gamma _{t})},\qquad  t
= 0,1.\nonumber
\end{eqnarray}
The goodness of fit of the model is tested in Section \ref{sec4.6}.

\begin{table}[t]
\tablewidth=205pt
\caption{Estimates and SE (in parenthesis) of model coefficients}
\label{tab3}
\begin{tabular*}{\tablewidth}{@{\extracolsep{\fill}}ld{2.8}d{2.8}@{}}
\hline
\textbf{Variable} & \multicolumn{1}{c}{\textbf{Private schools}} & \multicolumn{1}{c@{}}{\textbf{Public schools}}\\
\hline
\multicolumn{3}{@{}c@{}}{Assignment (logistic) model}\\
Const  & - 2.95\ (1.30) & 13.88\ (2.90)\\
$\delta_{t}$ & 0.49\ (0.21) & - 2.02\ (0.39)\\
GEN & 0.77\ (0.13) & - 0.76\ (0.18)\\
SEI & - 0.12\ (0.07) & 0.40\ (0.12)\\
HER & 3.16\ (0.20) & - 2.57\ (0.30)\\
SES & 0.09\ (0.07) & 0.27\ (0.11)\\
S.loc & 1.13\ (0.13) & - 1.63\ (0.24)\\[3pt]
\multicolumn{3}{@{}c@{}}{Population (normal) model}\\
$\sigma_{t}$ & 0.83\ (0.02) & 1.10\ (0.07)\\
Const & 6.09\ (0.07) & 6.89\ (0.14)\\
GEN & - 0.20\ (0.05) & 0.17\ (0.08)\\
ME & 0.18\ (0.05) & 0.11\ (0.07)\\
SEI & 0.16\ (0.03) & 0.16\ (0.04)\\
HER & 0.39\ (0.09) & 1.35\ (0.20)\\
SES & 0.21\ (0.02) & 0.30\ (0.04)\\
\hline
\end{tabular*}
\end{table}

When fitting the models to the two types of schools we found that the
x-variables contain all the variables listed in Section \ref{sec3.4} except for
school location (S.loc), which was found to be highly insignificant in
both the public and private school models. The v-variables contain all
the variables listed in Section \ref{sec3.4} except for mother's education (ME),
which was found to be highly insignificant in both the public and
private school models. As discussed in Section \ref{sec5}, the sample
\textit{pdf} (\ref{eq4.1}) obtained in the normal/logistic case is identifiable
and accommodates consistent and asymptotically normal (CAN) estimators
for all the model parameters, if $\mathrm{x}$ has at least one covariate not included in $\mathrm{v}$.

\subsubsection*{Results}

We computed the \textit{mle} of the unknown parameters by maximizing the
likelihood (\ref{eq4.3}) with respect to $\theta_{t} =
(\beta_{0t},\beta_{t},\sigma_{t}^{2})$ and $\alpha_{t} =
(\gamma_{0t},\delta_{t},\gamma_{t})$, $t = 0,1$.\vadjust{\eject} See Supplement B in the
supplementary material [\citet{PfeLan}] for the
maximization procedure. Table \ref{tab3} shows the estimates and standard errors
(SE) of the model coefficients.

As anticipated, $\hat{\delta} _{1} > 0$ and $\hat{\delta} _{0} < 0$, but
$\hat{\delta} _{1}$ is close to zero, although significant at the 5\%
level using the conventional $t$-statistic. On the other hand,
$\hat{\delta} _{0}$ is highly negative, indicating that for given values
of the covariates, the probability to attend a public school decreases
very rapidly as the proficiency score increases. This finding suggests
that pupils attending public schools are generally less able, and their  lower scores
are
not necessarily because of poor quality of the public
schools. Another important result emerging from Table \ref{tab3} is that the
variable school location (S.loc)
is highly significant in the assignment models even when including the response among the covariates.
We return to this
result in Section \ref{sec4.7}. The coefficient of S.loc is positive for private
schools and negative for public schools, suggesting that pupils from
urban areas tend to enroll in private schools.

Table \ref{tab4} shows the estimates of the population means by type of school
and the corresponding estimates of the ATE. We present the two estimates
defined in Section \ref{sec4.3}: the estimate $\hat{\mu} _{t,S}$ [equation (\ref{eq4.6})]
and the combined estimate $\hat{\mu} _{t,C}$ [equation (\ref{eq4.7})].

\begin{table}
\caption{Estimation of population means and ATE}
\label{tab4}
\vspace*{-3pt}
\begin{tabular}{@{}lccccd{2.2}d{2.2}@{}}
\hline
 & \multicolumn{2}{c}{\textbf{Private school}} & \multicolumn{2}{c}{\textbf{Public school}} & \multicolumn{2}{c@{}}{\textbf{ATE}}\\[-5pt]
 &\multicolumn{2}{c}{\hrulefill} &\multicolumn{2}{c}{\hrulefill}
 &\multicolumn{2}{c@{}}{\hrulefill}\\
 & $\bolds{\hat{\mu} _{1,S}}$ & $\bolds{\hat{\mu} _{1,C}}$ & $\bolds{\hat{\mu} _{0,S}}$ & $\bolds{\hat{\mu} _{0,C}}$ &
 \multicolumn{1}{c}{$\bolds{\hat{\tau} _{S} = \hat{\mu} _{1,S} - \hat{\mu} _{0,S}}$} &
 \multicolumn{1}{c@{}}{$\bolds{\hat{\tau} _{C} = \hat{\mu} _{1,C} - \hat{\mu} _{0,C}}$}\\
\hline
Estimate  & 6.10 & 6.09 & 7.05 & 6.91 & - 0.95 & - 0.82\\
Std. error & 0.05 & 0.06 & 0.15 & 0.12 & 0.16 & 0.13\\
\hline
\end{tabular}
\vspace*{-3pt}
\end{table}

\begin{table}[b]
\vspace*{-3pt}
\caption{Unweighted and weighted estimators of schools mean score and ATE}
\label{tab5}
\vspace*{-3pt}
\begin{tabular}{@{}lcccccc@{}}
\hline
\textbf{Method} & \multicolumn{2}{c}{\textbf{Private schools}} & \multicolumn{2}{c}{\textbf{Public schools}} & \multicolumn{2}{c@{}}{\textbf{ATE}}\\[-5pt]
&\multicolumn{2}{c}{\hrulefill}&\multicolumn{2}{c}{\hrulefill}&\multicolumn{2}{c@{}}{\hrulefill}\\
 & \textbf{UNWEI} & \textbf{WEI} & \textbf{UNWEI} & \textbf{WEI} & \textbf{UNWEI} & \textbf{WEI}\\
 \hline
Est. pop. regression  & 6.10 & 6.09 & 7.05 & 6.92 & $-$0.95 & $-$0.83\\
Combined estimator & 6.09 & 6.08 & 6.91 & 6.80 & $-$0.82 & $-$0.72\\
\hline
\end{tabular}
\end{table}

The two methods of estimating the population means yield similar
estimates and the ATE estimates are therefore likewise similar, negative
and highly significant, indicating the very interesting and important
result that after accounting for the school selection by pupils, the
mean score in the public schools is actually higher than in the private
schools. A similar conclusion was reached in Section \ref{sec3.4} by use of the
LV and IV methods, although with smaller ATE estimates.

\subsection{Probability weighted estimation under the proposed approach}\label{sec4.5}

We computed the PWE of the population means and the ATE under the
proposed approach, accounting for possible sampling effects, similarly
to the estimators computed under the previous methods shown in Section
\ref{sec3.5}. See Supplement A of the supplementary material [\citet{PfeLan}] for the corresponding expressions. Table \ref{tab5} shows the
unweighted (UNWEI) and probability weighted estimators (WEI) of the
school mean scores and the ATE. The results in this table reaffirm the
results in Section~\ref{sec3.5} that the PISA sampling design is not informative
for the models used for estimation of the ATE. We ignore the sample
selection process in subsequent analysis.

\subsection{Goodness of fit of the model fitted to the PISA data for Ireland}\label{sec4.6}

Assessing the goodness of fit of the model (\ref{eq4.8}), or, more generally,
any other model of the form (\ref{eq4.1}) seems formidable on first sight since
no observations are available from the population distribution under
strong ignorability, and the assignment probabilities are generally
unknown. Note, however, that once the identifiability of the sample
\textit{pdf} has been established (see Section~\ref{sec5}), one faces the
classical problem of having a random sample from a~hypothesized
\textit{pdf} that needs to be tested. Below we consider three tests,
which compare the theoretical and empirical sample distributions of the
responses and apply them to the PISA data. The power of the tests is
studied further in Appendix \ref{appB} by using simulated data sets.\vadjust{\eject}

\subsubsection*{Goodness-of-fit tests}

Suppose first that the true model parameters $\{ \alpha_{t},\break
\theta_{t}\}$ are known. Denote by $U_{t  i}(y) = F_{t  i}(y|\mathrm{z}_{i}) =
\int_{ - \infty} ^{y} f_{S_{t}}(y_{t  i}|\mathrm{z}_{i}; \alpha_{t}, \theta_{t})\,dy_{t  i}$ the
hypothesized \textit{cdf} of $Y_{t  i}|\mathrm{z}_{i},i =
1,2,\ldots,n_{t}$. For continuous $F_{t  i}$, the random variables
$U_{t  i}(y)$ evaluated at the outcomes $y_{t  i}$
are independent $\operatorname{Uniform} [0,1]$ variables since the responses $Y_{t  i}$ are independent given the covariates $\mathrm{z}_{i}$. Let
$u_{t1},\ldots,u_{t  n_{t}}$ denote the values of
$U_{t1},\ldots,U_{t  n_{t}}$ computed at the sample values
$y_{t1},\ldots,y_{t  n_{t}}$ and let $F_{t,\mathrm{EMP}}$ define
the empirical distribution of $u_{t1},\ldots,u_{t  n_{t}}$. The
following goodness-of-fit tests are in common use, where
$u_{t(1)},\ldots,u_{t(n_{t})}$ are the ordered values of
$u_{t1},\ldots,u_{t  n_{t}}$ [\citet{Ste86}]:
\begin{eqnarray}\label{eq4.9}
\qquad &&\mbox{Kolmogorov--Smirnov:}\nonumber\\[-8pt]\\[-8pt]
\qquad &&\qquad \mathit{KS}_{t} = \max_
{i}\bigl|F_{t,\mathrm{EMP}}\bigl(u_{t(i)}\bigr) - u_{t(i)}\bigr|,\nonumber \\
\label{eq4.10}
\qquad &&\mbox{Cramer--von Misses:}\nonumber\\[-8pt]\\[-8pt]
\qquad &&\qquad \mathit{CM}_{t} = \frac{1}{12n_{t}} + \sum_{i =
1}^{n_{t}} \biggl[ u_{t(i)} - \frac{2i - 1}{2n_{t}} \biggr] ^{2},\nonumber \\
\label{eq4.11}
\qquad &&\mbox{Anderson--Darling:}\nonumber\\[-8pt]\\[-8pt]
\qquad &&\qquad \mathit{AD}_{t} = - n_{t} - \frac{1}{n_{t}}\sum_{i =
1}^{n_{t}} \bigl[(2i - 1)\ln\bigl(u_{t(i)}\bigr) + (2n_{t} + 1 - 2i)\ln\bigl(1 -
u_{t(i)}\bigr)\bigr].\nonumber
\end{eqnarray}

It is known [e.g., \citet{BabFei06}] that KS is sensitive to
large-scale differences in location and shape between the model and the
empirical distribution, CM is sensitive to small-scale differences in
the shape and AD is sensitive to differences near the tails of the
distribution. All the three test statistics are
\textit{distribution-free} as long as the hypothetical \textit{cdf} is
fully specified (known parameters).

When computed with estimated parameters, the asymptotic distribution of
the three statistics depends in a complex way on the true underlying
\textit{cdf}, and possibly also on the method of estimation. Correct
critical values can be obtained in this case by use of parametric
bootstrap. The procedure consists of generating a large number of
samples from the estimated hypothesized model, re-estimating the unknown
model parameters from each bootstrap sample and then computing the
corresponding test statistics. The bootstrap distribution of the test
statistics approximates the true distributions under the hypothesized
model with correct order of error [Babu and Rao (\citeyear{JogRao04})].

\subsubsection*{Validating the model fitted to the PISA data for Ireland}

As explained above, the critical values for the distribution of the test
statistics can be computed from the bootstrap distribution. To this end,
we generated 250 bootstrap samples for each type of school $(t = 0, 1)$
by first generating new outcomes $Y_{t  i}$ from the estimated
normal population model $f_{p}(y_{t
i}|\mathrm{x}_{i};\hat{\beta} _{t},\hat{\sigma} _{t}^{2})$,\vadjust{\eject} using the
same covariates as for the actual sample, and then selecting the units
to the sample $S_{t}$ using the estimated logistic
probabilities $\Pr(T_{i} = t|\mathrm{v}_{i},y_{t
i};\hat{\delta} _{t},\hat{\gamma} _{t})$. Notice that this way the
sample sizes in the two groups are no longer constant. We found that for
the treatment group the mean sample size is 1,245 with standard deviation
17, and for the control group the mean sample size is 700 with standard
deviation 18. (The sample sizes in the true data set are 1,244 and 694,
resp.) Next we computed the \textit{mle} of the parameters
$\beta_{t},\sigma_{t}^{2},\gamma_{t},\delta_{t}$ for each bootstrap
sample and the test statistics~(\ref{eq4.9})--(\ref{eq4.11}).

Table \ref{tab6} shows the values of the three test statistics for the PISA
samples and their $p$-values, as computed from the corresponding bootstrap
distributions. As can be seen, all the three statistics are
nonsignificant with $p$-values higher than 10\%, thus supporting the use
of the selected models.

\begin{table}[t]
\caption{Goodness-of-fit test statistics and $p$-values (in parenthesis)}
\label{tab6}
\begin{tabular}{@{}lccccc@{}}
\hline
\multicolumn{3}{@{}c}{\textbf{Private schools}} & \multicolumn{3}{c@{}}{\textbf{Public schools}}\\[-5pt]
\multicolumn{3}{@{}c}{\hrulefill}&\multicolumn{3}{c@{}}{\hrulefill}\\
\textbf{KS} & \textbf{CM} & \textbf{AD} & \textbf{KS} & \textbf{CM} & \textbf{AD}\\
\hline
0.023 (0.12) & 0.089 (0.18) & 0.62 (0.11) & 0.027 (0.17) & 0.062 (0.32) & 0.45 (0.15)\\
\hline
\end{tabular}
\end{table}

\begin{remark}\label{rem8}
We also computed the
empirical means and standard deviations over the 250 bootstrap samples
of the estimates of the model coefficients and all the ATE estimates
considered in Sections \ref{sec3} and \ref{sec4}. To save in space, we do not show the
detailed results, but the means are generally close (and in most cases
very close) to the values computed for the PISA data, and likewise for
the standard deviations. These results indicate that the model
coefficients can be estimated almost unbiasedly and with acceptable
standard error estimates, despite the complicated structure of the
sample model. Obtaining similar ATE estimates under the different
methods to the estimates computed from the PISA data is another
indication of the goodness of fit of the models fitted to this data
set.
\end{remark}

\subsection{Testing the assumptions underlying existing methods}\label{sec4.7}

As mentioned earlier, the use of the proposed approach enables testing
some of the assumptions underlying existing methods under the sample
model fitted to the observed data. Consider first the logistic models
for the assignment probabilities. The coefficients~$\hat{\delta} _{t}$
of the response are significant in both models, with~$\hat{\delta} _{0}$,
in particular, being highly negative. This result suggests that the
covariates available for the present study are not sufficient to explain
the choice of school, and, hence, that methods that assume strong
ignorability [assumptions $\operatorname{SI}(a)$--$\operatorname{SI}(b)$] and,
in particular, methods that employ
propensity scores computed with these covariates are not adequate.
Notice\vadjust{\eject} in Table~\ref{tab1} that the use of these methods yields positive ATE
estimates, although of lower magnitude than the crude
difference, $\bar{y}_{1} - \bar{y}_{0}$.

Next consider the IV method. In its simple form it assumes the
model $Y_{t} = \mathrm{x}'\beta_{t} + u_{t};E_{p}(u_{t}|\mathrm{x}) = 0$,
with three added conditions on the instrument $h$:
IV(a)---$E_{p}(Y_{t}|\mathrm{x},h) = E_{p}(Y_{t}|\mathrm{x})$,
IV(b)---$E_{p}[T(u_{1} - u_{0})|\mathrm{x},h] = 0$, and IV(c)---$\Pr(T =
1|\mathrm{x},h)\ne\Pr(T = 1|\mathrm{x})$. The population model with the
covariates $\mathrm{x}$ listed in Section~\ref{sec3.4} is validated in Section
\ref{sec4.6}. Furthermore, our analysis shows that the instrument S.loc is highly
insignificant in the two population models, thus supporting the
condition IV(a). Condition IV(b) cannot be verified empirically, but
this condition is generally considered as being mild and it can be
relaxed further [\citet{Woo02}]. Finally, the condition~IV(c) is
verified as well since the coefficients of the instrument in the models
fitted for the assignment probabilities are highly significant (Table
\ref{tab3}), even when including the response $y$ as an additional explanatory
variable. Indeed, the use of the IV method yields an ATE estimate of
$-0.61$ (Table~\ref{tab1}), which is the closest to the estimate obtained under our
approach among the other methods considered.\vspace*{-2pt}

\begin{remark}\label{rem9}
We emphasize again that all the above conclusions are
under the model that we have fitted (and validated) to the data.\vspace*{-1pt}
\end{remark}

\section{Foundation of proposed approach}\label{sec5}\vspace*{-2pt}

\subsection{Identification of the sample distribution}\label{sec5.1}

An important question underlying the use of the sample \textit{pdf}
(\ref{eq4.1}) is model identification. In order to get some insight into this
issue, we restrict to a given treatment $t$ and hence drop for
convenience the subscript $t$ everywhere, denoting by
$p(y,\mathrm{v};\alpha) = \Pr(i \in S_{t}|y_{t},\mathrm{v};\alpha)$ the
probability assignment rule (PAR) to the sample $S_{t}$, and by
$f_{p}(y|\mathrm{x};\theta)$ the corresponding population \textit{pdf}
of $Y_{t}|\mathrm{x}$ under strong ignorability. We assume that the
response is continuous. Let $J \subseteq R$ define the common domain of
the y-values for these functions. The sample \textit{pdf} for units in
$S_{t}$ is therefore $f_{S_{t}}(y|\mathrm{x},\mathrm{v};\theta,\alpha) =
\frac{f_{p}(y|\mathrm{x};\theta )p(y,\mathrm{v};\alpha )}{\int
f_{p}(y|\mathrm{x};\theta )p(y,\mathrm{v};\alpha )\,dy}$ and the
identifiability of the sample \textit{pdf} is defined as follows:\vspace*{-2pt}

\begin{definition}\label{defin1}
The sample \textit{pdf}
$f_{S_{t}}(y|\mathrm{x},\mathrm{v};\theta,\alpha)$ is identifiable if no
different (proper) densities $f_{p}^{(1)}(y|\mathrm{x};\theta^{(1)}),
f_{p}^{(2)}(y|\mathrm{x};\theta^{(2)})$ and different
PARs $p^{(1)}(y,\break\mathrm{v};\alpha^{(1)}),
p^{(2)}(y,\mathrm{v};\alpha^{(2)})$ exist such that the
pairs
$[f_{p}^{(1)}(y|\mathrm{x};\theta^{(1)}),p^{(1)}(y,\mathrm{v};\alpha^{(1)})]$
and
$[f_{p}^{(2)}(y|\mathrm{x};\theta^{(2)}),p^{(2)}(y,\mathrm{v};\alpha^{(2)})]$
induce the same sample \textit{pdf} for every $y \in J$ and every set of
covariates $\mathrm{(x,v)}$.\vspace*{-2pt}
\end{definition}

Clearly, if different pairs $[f_{p}^{(1)},p^{(1)}],
[f_{p}^{(2)},p^{(2)}]$ yield the same sample\break \textit{pdf},~the~model is
not identifiable. At first thought, this would seem to be always the
case since (\ref{eq4.1}) is the same \textit{pdf} for
the
pair $[f_{p}^{(1)}(y|\mathrm{x};\theta^{(1)}),\break p^{(1)}(y,\mathrm{v};\alpha^{(1)})]$,
and when the population \textit{pdf} is\vadjust{\eject}
$f_{p}^{(2)}(y|\mathrm{x},\mathrm{v};\theta^{(1)},\alpha^{(1)}) =\break
\frac{f_{p}^{(1)}(y|\mathrm{x;}\theta ^{(1)})p^{(1)}(y,\mathrm{v};\alpha
^{(1)})}{\int f_{p}^{(1)}(y|\mathrm{x;}\theta
^{(1)})p^{(1)}(y,\mathrm{v};\alpha ^{(1)})\,dy}$ and the units are
assigned with probabilities that do not depend on $y$ given $\mathrm{v}
$ (ignorable assignment). The
\textit{pdf} $f_{p}^{(2)}(y|\mathrm{x},\mathrm{v};\theta^{(1)},\break\alpha^{(1)})$,
however, does not generally belong to a conventional parametric family
and is very different from $f_{p}^{(1)}(y|\mathrm{x};\theta^{(1)})$,
especially when the assignment mechanism is strongly informative. Hence,
field experts should be able to decide which of the two \textit{pdf}s,
$f_{p}^{(1)}(y|\mathrm{x};\theta^{(1)})$ or
$f_{p}^{(2)}(y|\mathrm{x},\mathrm{v};\theta^{(1)},\alpha^{(1)})$,
is a more sensible population \textit{pdf} for the potential responses
in a given problem.

\subsubsection*{Conditions for model identification}

Lemma \ref{lem1} defines different conditions under which the sample \textit{pdf}
is identifiable. We assume for convenience that there are no covariates,
but all the results generally hold when covariates~$\mathrm{x,v}$ exist.
Define $R_{p}(y;\theta^{(1)},\theta^{(2)}) = \frac{f_{p}^{(2)}(y;\theta
^{(2)})}{f_{p}^{(1)}(y;\theta ^{(1)})}$. We assume throughout this
section that the functions $f^{(j)}_{p}(y;\theta^{(j)})$
and $p^{(j)}(y;\alpha^{(j)}), j = 1,2$, are strictly positive on $J^{*}
\subseteq J$.

\begin{lemma}\label{lem1}
\textup{(a)} Suppose that $J^{*} = [c,\infty)$ for some constant $c$. If
$f_{p}^{(1)}(y;\break\theta^{(1)})$ and $f_{p}^{(2)}(y;\theta^{(2)})$ are two
different \textit{pdf}s satisfying for any
given $\theta^{(1)},\theta^{(2)},\break\lim_{y \to \infty}
R_{p}(y;\theta^{(1)},\theta^{(2)}) = 0,\infty$ or does not exist, then
there are no different PARs $p^{(1)}(y;\alpha^{(1)}),
p^{(2)}(y;\alpha^{(2)})$ with finite positive limits as $y \to\infty$
yielding the same sample pdf for all $y \in J^{*}$.

\textup{(b)} Suppose that $J^{*} = ( - \infty,c]$ for some constant $c$. If
$f_{p}^{(1)}(y;\theta^{(1)})$ and $f_{p}^{(2)}(y;\theta^{(2)})$ are two
different \textit{pdf}s satisfying for any
given $\theta^{(1)},\theta^{(2)},\break\lim _{y \to - \infty}
R_{p}(y;\theta^{(1)},\theta^{(2)}) = 0,\infty$ or does not exist, then
there are no different PARs $p^{(1)}(y;\alpha^{(1)}),
p^{(2)}(y;\alpha^{(2)})$ with finite positive limits as $y \to - \infty$
yielding the same sample \textit{pdf} for all $y \in J^{*}$.

\textup{(c)} Let $y_{0}$ be a limit point of $J^{*}$. If
$f_{p}^{(1)}(y;\theta^{(1)})$ and $f_{p}^{(2)}(y;\theta^{(2)})$ are two
different \textit{pdf}s satisfying for any
given $\theta^{(1)},\theta^{(2)}, \lim _{y \to y_{0}^{ +}
(y \to y_{0}^{ -} )}R_{p}(y;\theta^{(1)},\theta^{(2)}) = 0,\infty$ or
does not exist, then there are no different PARs
$p^{(1)}(y;\alpha^{(1)}),p^{(2)}(y;\break\alpha^{(2)})$ with finite positive
limits at $y = y_{0}$ yielding the same sample \textit{pdf} for all $y
\in J^{*}$.
\end{lemma}

\begin{pf}
Part (a) is similar to \citet{LeeBer01} and is
proved in Supplement C of the supplementary material [\citet{PfeLan}].
The proofs of the other two parts are similar.
\end{pf}

Lemma \ref{lem1} enables verifying the identifiability of the sample \textit{pdf}
for many combinations of population \textit{pdf}s and PARs, but there
are cases that need to be studied separately. For example, the lemma is
not applicable to the case of normal population \textit{pdf}s and
logistic PARS if the coefficients of $y$ in the two logistic
distributions are allowed to have opposite signs.\vadjust{\eject} This is because in
this case one of the PARS will have a limit of zero as $y \to\infty$
or $y \to - \infty$, and the other PAR will have a limit of 1 (but see
Result \ref{res1} below).

\subsubsection*{Further model identification results}

Result \ref{res1} states the identifiability of the sample \textit{pdf} resulting
from the combination of a normal population \textit{pdf} and a logistic
\textit{PAR}. The proof is given in Supplement D of the supplementary
material [\citet{PfeLan}]. \citet{Lan} considers
other combinations of population \textit{pdf}s and \textit{PAR}s.

\begin{result}\label{res1}
No different pairs
$[f_{p}(y|\mathrm{x};\theta), p(y,\mathrm{v};\alpha)]$ of a normal
\textit{pdf} and logistic \textit{PAR} yield the same sample
\textit{pdf}, if the vectors $\mathrm{x}$ and v differ in at least one
covariate.
\end{result}

\begin{remark}\label{rem10}
The condition on the covariates seems to impose a
limitation on the model, but, in practice, there is no reason why the
covariates used to model the response under strong ignorability should
be the same as the covariates used to model the treatment assignment
probabilities. See the models in Section \ref{sec4.4}.
\end{remark}

\subsection{Practical identifiability}\label{sec5.2}

Section \ref{sec5.1} and the additional results in Lands\-man (\citeyear{Lan}) establish the
``theoretical identifiability'' of the sample model under a large number
of plausible combinations of population \textit{pdf}s and \textit{PAR}s.
It is important to mention, however, that identifiability problems may
arise in practice, depending on the forms of
$f_{p}(y|\mathrm{x};\theta)$ and $p(y,\mathrm{v};\alpha)$. For example,
\citet{LeeBer01} consider the case $f_{p}^{(1)}(y) =
N(0,1),p^{(1)}(y) = \Phi(y - 1)$. The authors show graphically that in
this case the sample density $f_{S}(y) = f_{p}^{(1)}(y)p^{(1)}(y)/\int
f_{p}^{(1)}(y)p^{(1)}(y)\,dy$ can be closely approximated by the normal
density $N(0.92,0.79^{2})$. This means that even though the sample
\textit{pdf} $f_{S}(y)$ is theoretically identifiable [\citet{Lan}],
a problem may arise in practice when fitting the model, distinguishing
between this density and the sample density obtained from $f_{p}^{(2)}(y)
= N(0.92,0.79^{2}),p^{(2)}(y) = \mathrm{const}$. \citet{LeeBer01} refer to
this phenomenon as ``practical nonidentifiability.'' Another example is\vspace*{1pt}
where the \textit{PAR} is logistic. Suppose that $p^{(1)}(y) = \{ 1 +
[\exp(1 + y)]^{ - 1}\} ^{ - 1},p^{(2)}(y) = \{ 1 + [\exp( - 1 - y)]^{ -
1}\} ^{ - 1}$. Then, the pairs $[f_{p}^{(1)}(y) = N(0,1), p^{(1)}(y)]$
and $[f_{p}^{(2)}(y) = N(1,1), p^{(2)}(y)]$ induce the same sample
\textit{pdf} (``theoretical nonidentifiability''), and this \textit{pdf}
is closely approximated by $N(0.28,0.93^{2})$ (``practical
nonidentifiability'').

We emphasize that in the presence of covariates, if the vectors x and v
differ in at least one covariate, the problem of practical
identifiability will generally not exist. See \citet{Lan} for a
detailed analysis.

\subsection{Asymptotic properties of the maximum likelihood estimators}\label{sec5.3}

The \textit{mle} $\{ \hat{\alpha} _{t} , \hat{\theta} _{t} , t = 0,1\}$
obtained by maximization of (\ref{eq4.3}) are the solutions\vadjust{\eject} of the estimating
equations $\sum_{k = 1}^{n} u_{t,k} = 0$, where $u_{t,k}$ is the vector
of the first derivatives of the $k$th component of the log-likelihood
with respect to $(\alpha_{t} , \theta_{t})$. \citet{RotRob97}
show that no $\sqrt{n}$-consistent and asymptotically normal (CAN)
estimator of $\theta_{t}$ exists if (and only if) the derivatives of the
log-likelihood with respect to $\theta_{t}$ are collinear with the
derivatives of the log-likelihood with respect to $\alpha_{t}$ with
probability 1. The derivatives are evaluated at the true parameter
values. The authors illustrate that if the population model is
$Y_{ti}\sim N(\beta_{t},1)$ and $\Pr(T_{i} = t|y_{ti}) = \frac{\exp
(\gamma _{0t} + \delta _{t}y_{t  i})}{1 + \exp (\gamma _{0t} +
\delta _{t}y_{t  i})}$, then no CAN estimator for $\beta_{t}$
exists if in truth $\delta_{t} = 0$ (ignorable assignment).\looseness=1

\citet{Lan} shows
that if $f_{p}(y_{t  i}|\mathrm{x}_{i}) = N(\beta_{0t} +
\mathrm{x}'_{i}\beta_{t},\sigma_{t}^{2}), \mathrm{logit}[\Pr(T_{i} =
t|y_{t  i},\mathrm{v}_{i})]= \gamma_{0t} +
\delta_{t}y_{t  i} + \mathrm{v}'_{i}\gamma_{t}$ and
x has at least one covariate not included in~v, the derivatives
of the log-likelihood (\ref{eq4.3}) with respect to
$\theta_{t} = (\beta_{0t},\beta_{t},\sigma_{t})$ are not collinear with
the derivatives with respect to $\alpha_{t} =
(\gamma_{0t},\gamma_{t},\delta_{t})$ with probability 1, and, hence, CAN
estimators for the parameters exist even when the true assignment
process is ignorable. This property enables testing the ignorability of
the assignment process, as illustrated in Sections \ref{sec4.4}
and~\ref{sec4.7}.\looseness=1

\section{Summary remarks}\label{sec6}

In this article we study the use of an alternative approach for
observational studies that recovers the treatment assignment model and
the population model under strong ignorability from the sample data. It
is shown that the sample model holding for the observed data, which
incorporates the two models, is identifiable under some general
conditions. Furthermore, the goodness of fit of the sample model can be
tested successfully by standard test statistics because the sample model
refers to the observed data. As illustrated in Section \ref{sec4.7}, the sample
model enables also testing the appropriateness of the use of some of the
existing methods for a particular data set.

We applied the new approach for comparing the proficiency in mathematics
of children aged 15 between public and private schools in Ireland. Our
analysis shows that although the average score of pupils in the sample
from private schools is significantly higher than the average score of
pupils from public schools, the picture is reversed once the effect of
the school selection is accounted for properly. A similar conclusion is
reached by the methods of latent variables and instrumental variables.

The approach proposed in this article is fully parametric, which raises
questions of its robustness to departures from the models fitted to the
data. We emphasize again that the models can be tested and, as the
empirical illustrations show, the test statistics that we have applied
have good power properties. Nonetheless, it is certainly worth
considering the use of semi-parametric or nonparametric models either
for the population models under strong ignorability and/or for the
assignment probabilities, thus further robustifying the inference.

\begin{appendix}
\section{Imputation of proficiencies in the PISA study}\label{appA}\vspace*{-2pt}

\renewcommand{\theequation}{A.1}

The multiple imputation approach draws at random multiple values from
the conditional distribution of the unobserved proficiency, $\psi_{j}$ of
pupil $j$, given the $m$ observed responses $d_{j} =
(d_{1j},\ldots,d_{m j})$ and covariates $\mathrm{x}_{j}$
representing individual background characteristics. The conditional
\textit{pdf} of $\psi_{j}$ is expressed
as
\begin{eqnarray}
f(\psi_{j}|d_{j},\mathrm{x}_{j}) &\propto&\prod_{i = 1}^{m} [\Pr(D_{ij}
= 1|a_{i},b_{i},\psi_{j})]^{d_{ij}} \nonumber\\[-10pt]\\[-10pt]
&&\hphantom{\prod_{i = 1}^{m}}{}\times[\Pr(D_{ij} = 0|a_{i},b_{i},\psi_{j})]^{(1 -
d_{ij})}f(\psi_{j}|\mathrm{x}_{j},\lambda,\sigma),\nonumber
\end{eqnarray}
where $f(\psi_{j}|\mathrm{x}_{j},\lambda,\sigma^{2})$ is the normal
distribution with mean $\mathrm{x}'_{j}\lambda$ and
variance~$\sigma^{2}$, and $\Pr(D_{ij} = 1|a_{i},b_{i},\psi_{j})= [1 +
\exp( - a_{i}(\psi_{j} - b_{i}))]^{ - 1}$. The parameter $a_{i}$ measures
how question $i$ distinguishes between pupils and the parameter $b_{i}$
represents the ``difficulty'' of question $i$. The responses to the various
questions are assumed to be independent given $(a_{i},b_{i},\psi_{j})$.
Five imputed values of $\psi_{j}$ are drawn for every student $j$ in
the sample and stored in the PISA database.\vspace*{-2pt}

\section{Powers of goodness-of-it test statistics}\label{appB}\vspace*{-2pt}

In Section \ref{sec4.6} we considered three goodness-of-fit test statistics and
applied them for testing the model (\ref{eq4.8}) fitted to the PISA data. In
order to study the powers of the three test statistics in the case of
misspecified population models, we simulated new data sets for each of
the two groups (public and private schools) from the same models as
in Section \ref{sec4.6}, except that the residual terms in the two population models were
generated from the skew $t$-distribution defined by \citet{AzzCap03}. The true population means and hence the ATE remain
unchanged. The skew $t$-distribution depends on four parameters:
location ($\xi)$, scale ($w)$, shape~($\alpha)$ and degrees of freedom
($\upsilon)$. The normal and $t$-distributions are members of this
family of distributions. For example, the case $\xi = 0$, $w = 1$, $\alpha =
0$, $\upsilon = \infty$ defines the standard normal distribution.

\begin{figure}

\includegraphics{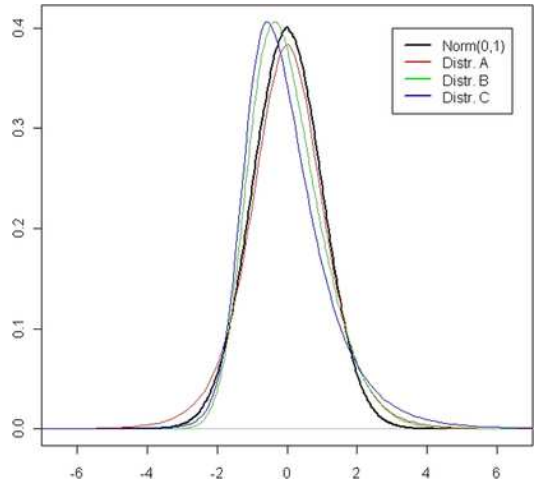}

  \caption{Densities of residuals under the four alternative distributions.}\label{fig1}
\end{figure}

We generated 100 sets of residuals for each group from the following 3
distributions: (A) $\xi = 0$, $w = 1$, $\alpha = 0$, $\upsilon = 6$, (B) $\xi =
- 1.16$, $w = 1.55$, $\alpha = 2.5$, $\upsilon = \infty$, (C) $\xi = - 1.24$, $w
= 1.45$, $\alpha = 2.5$, $\upsilon = 6$. Distribution A defines the standard
$t$-distribution with 6 degrees of freedom. Distribution B defines a
skewed distribution with relatively short tails, while Distribution C
defines a skewed distribution with a heavy right tail. The three
densities are plotted in Figure \ref{fig1}. The location parameters were chosen
in such a way that all the three distributions have mean zero, implying
that the true ATEs are the same. The standard deviations equal 1.22,
1.04 and 1.27, respectively. Next we fitted the models that assume that
the residuals are normal [equation~(\ref{eq4.8})], such that the true models are
misspecified.\vadjust{\eject}

\renewcommand{\thetable}{B1}
\begin{table}
\tabcolsep=0pt
\caption{Average estimates (SE) and percent of samples with rejected model}
\label{tabB1}
\hspace*{-10pt}
\begin{tabular*}{1.05\tablewidth}{@{\extracolsep{4in minus 4in}}lc@{\hspace*{20pt}}c@{\hspace*{20pt}}c@{\hspace*{20pt}}cc@{\hspace*{20pt}}c@{\hspace*{20pt}}c@{\hspace*{20pt}}c@{}}
\hline
 & \multicolumn{4}{c}{\textbf{Private schools}} & \multicolumn{4}{c@{}}{\textbf{Public schools}}\\[-5pt]
 & \multicolumn{4}{c}{\hrulefill} & \multicolumn{4}{c@{}}{\hrulefill}\\
 & \multicolumn{4}{c}{$\bolds{\hat{\mu} _{1,S} = 6.01\, (0.14),\hat{\mu} _{1,C} = 5.99\,  (0.18)}$} &
   \multicolumn{4}{c}{$\bolds{\hat{\mu} _{0,S} = 7.34\, (0.13),\hat{\mu} _{0,C} = 7.09\, (0.39)}$}\\[-5pt]
 & \multicolumn{4}{c}{\hrulefill} & \multicolumn{4}{c@{}}{\hrulefill}\\
\textbf{ATE} & \multicolumn{8}{c@{}}{$\bolds{\hat{\tau} _{S} = - 1.32\,  (0.20); \hat{\tau} _{C} = - 1.10\, (0.43)}$}\\
\hline
\textbf{Sig. level} & \textbf{0.10} & \textbf{0.05} & \textbf{0.025} & \textbf{0.01} & \textbf{0.10} & \textbf{0.05} & \textbf{0.025} & \textbf{0.01}\\
\hline
\multicolumn{2}{@{}l}{Distribution A} \\
\quad K--S & 100 & 100 & \hphantom{0}99 & \hphantom{0}96 & 75 & 64 & 55 & 52\\
\quad C--M & 100 & 100 & 100 & 100 & 91 & 87 & 79 & 72\\
\quad A--D & 100 & 100 & 100 & 100 & 94 & 90 & 88 & 83\\[6pt]
&\multicolumn{4}{c}{$\bolds{\hat{\mu} _{1,S} = 5.96\,  (0.19),\hat{\mu} _{1,C} = 6.01\,  (0.15)}$} &
\multicolumn{4}{c}{$\bolds{\hat{\mu} _{0,S} = 6.84\,  (0.10),\hat{\mu} _{0,C} = 6.89\,  (0.26)}$}\\[-5pt]
 & \multicolumn{4}{c}{\hrulefill} & \multicolumn{4}{c@{}}{\hrulefill}\\
\textbf{ATE} & \multicolumn{8}{c@{}}{$\bolds{\hat{\tau} _{S} = - 0.88\,  (0.22) ; \hat{\tau} _{C} = - 0.88\,  (0.31)}$}\\
\hline
\textbf{Sig. level} & \textbf{0.10} & \textbf{0.05} & \textbf{0.025} & \textbf{0.01} & \textbf{0.10} & \textbf{0.05} & \textbf{0.025} & \textbf{0.01}\\
\hline
\multicolumn{2}{@{}l}{Distribution B} \\
\quad K--S & 100 & 100 & 100 & \hphantom{0}99 & 72 & 64 & 51 & 47\\
\quad C--M & 100 & 100 & 100 & 100 & 83 & 77 & 66 & 63\\
\quad A--D & 100 & 100 & 100 & 100 & 92 & 79 & 76 & 70\\[6pt]
 & \multicolumn{4}{c}{$\bolds{\hat{\mu} _{1,S} = 5.42\, (0.20), \hat{\mu} _{1,C} = 5.49\, (0.67)}$} &
 \multicolumn{4}{c}{$\bolds{\hat{\mu} _{0,S} = 6.76\, (0.08),\hat{\mu} _{0,C} = 6.90\, (0.46)}$}\\[-5pt]
 & \multicolumn{4}{c}{\hrulefill} & \multicolumn{4}{c@{}}{\hrulefill}\\
\textbf{ATE} & \multicolumn{8}{c@{}}{$\bolds{\hat{\tau} _{S} = - 1.34\, (0.22) ; \hat{\tau} _{C} = - 1.41\, (0.83)}$}\\
\hline
\textbf{Sig. level} & \textbf{0.10} & \textbf{0.05} & \textbf{0.025} & \textbf{0.01} & \textbf{0.10} & \textbf{0.05} & \textbf{0.025} & \textbf{0.01}\\
\hline
\multicolumn{2}{@{}l}{Distribution C}\\
\quad K--S & 100 & 100 & 100 & 100 & 67 & 56 & 40 & 39\\
\quad C--M & 100 & 100 & 100 & 100 & 85 & 75 & 71 & 67\\
\quad A--D & 100 & 100 & 100 & 100 & 93 & 88 & 79 & 78\\
\hline
\end{tabular*}
\end{table}

Table \ref{tabB1} shows the percentage of samples that the misspecified models
have been rejected by the three test statistics when using the
conventional 10\%, 5\%, 2.5\% and 1\% significance levels. The
percentages estimate the power of the tests. As can be seen, all the
tests reject the three misspecified models for the private schools in
literally all the samples, and the powers of the Cramer--von Misses test,
and, in particular, the powers of the Anderson--Darling test are
acceptable for the three misspecified models for the public schools as
well. We mention in this regard that the three misspecified
distributions were chosen to be sufficiently close to the normal
distribution (see Figure \ref{fig1}). Further empirical results not shown
indicate that for only mild larger distortions from the normal
distribution, the powers of the three tests for the public schools are
likewise close to 1.

Table \ref{tabB1} also shows the average of the estimates of the mean score in
each group and the ATE, and the corresponding standard errors of the
estimates (in parenthesis), over the 100 data sets. The empirical
averages deviate now significantly from the corresponding true values
($\mu^{1}=6.10$, $\mu^{0}=7.05$; $\Delta=-0.95$)
under all the three misspecified models and with larger standard errors
than under the correct model. However, the estimates of the ATE are
still highly negative, as under the correct model.

The conclusion from this simulation study is that even mild distortions
from normality can affect the magnitude of the estimates of the ATE (but
not their sign), but these mild distortions can be detected by the
goodness-of-fit test statistics.
\end{appendix}

\section*{Acknowledgments}
This paper is based on the Ph.D. thesis of the second
author written at the Hebrew University of Jerusalem, Israel, under the
supervision of the first author. The authors are grateful
to the Editor, Associate Editor and two referees for very insightful and
constructive comments that improved the quality of the paper.

\begin{supplement}[id=suppA]
\stitle{Supplement to: ``Are private schools better than public schools?
Appraisal for Ireland by methods for observational studies''\\}
\slink[doi]{10.1214/11-AOAS456SUPP}  
\slink[url]{http://lib.stat.cmu.edu/aoas/456/supplement.zip}
\sdatatype{.zip}
\sdescription{This supplement contains a PDF which is divided into five sections:
\begin{longlist}[Supplement A]
\item[Supplement A] develops the probability weighted estimators of the ATE.
\item[Supplement B] describes the maximization of the likelihood (\ref{eq4.3}).
\item[Supplement C] contains the proof of Lemma \ref{lem1}.
\item[Supplement D] contains the proof of Result \ref{res1}.
\item[Supplement E] describes the data file, which is provided.\\
\indent The data file PISA\_math2000.R contains the data.
\end{longlist}}
\end{supplement}


\printaddresses

\end{document}